\documentclass[a4paper,11pt]{article}
\pdfoutput=1 

\usepackage{jheppub} 

\usepackage[T1]{fontenc} 

\title{Gravitational energy and radiation of a charged black hole}

\author[a]{Luciano Combi,}
\author[a,b]{Gustavo E. Romero,}

\affiliation[a]{Instituto Argentino de Radioastronom\'ia (CCT-La Plata, CONICET; CICPBA), C.C. No. 5, 1894, Villa Elisa, Argentina}
\affiliation[b]{Facultad de Ciencias Astron\'omicas y Geof\'isicas, Universidad Nacional de La Plata, Paseo del Bosque s/n, 1900 La Plata, Buenos Aires, Argentina.}

\emailAdd{lcombi@iar.unlp.edu.ar}

\abstract{We analyze the energy configuration of a charged black hole in the Teleparallel Framework of General Relativity. We obtain the energy-momentum tensor of the gravitational field in a stationary frame, and we calculate its contribution to the total energy of the system. We study the same gravitational field measured by an accelerated frame and we analyze how the energy-momentum tensor is transformed. We found that in the accelerated frame, a Poyinting-like flux appears for the gravitational field but not for the electromagnetic field.}

\begin{document} 
\maketitle
\flushbottom

\section{\label{sec:level1} Introduction}

Most concepts of energy in physics are frame dependent. In Minkowski space-time, there is a preferred class of reference systems corresponding to the associated symmetries: the inertial-Cartesian frames. When space-time has no symmetries, defining a conserved property such as energy or momentum is problematic. The concept of gravitational energy is even more cumbersome. Since gravity effects vanish locally for a free falling frame, any well-behaved energy concept should be to some extent non-local and strongly frame dependent.

In recent years, the Teleparalell Framework of General Relativity (GR$_{||}$) proved to be a viable and interesting approach to deal with this problem \cite{Maluf:2013gaa}. The theory has the tetrad field as the dynamical object, instead of the metric, and posses equivalent field equations to those of General Relativity. However, contrary to General Relativity, it is possible to deduce from its field equations a conserved energy-momentum tensor for gravity. This energy-momentum tensor is covariant but depends globally on the tetrad field (for an analysis of the non-local character of this tensor see Ref. \cite{Combi}). This gauge dependence, typical of any definition of gravitational energy is physically meaningful in GR$_{||}$ when we interpret the tetrad field as a reference system \cite{Maluf:2007qq}.

Every quasi-local energy concept in GR is concerned with the total gravitational plus matter energy in a closed 2-surface \cite{lrr-2009-4}. This is also the case of GR$_{||}$, where the total conserved four momentum of the field can be casted as a surface integral using Gauss's theorem. Nevertheless, since the theory admits an energy-momentum tensor for each field, we can also analyze their contribution to the total energy separately. This can be useful for a better understanding of the interaction between matter and the gravitational field. In this paper, we will consider a simple non-vacuum scenario to analyze this problem: the Reissner-Nordstrom space-time.

\section{Teleparallel energy and frames}

The Teleparallel framework of General Relativity is an alternative theory of gravity whose dynamical object is the tetrad field $\mathbf{e}_{a}(x)= e_{a}^{\mu}(x) \partial_{\mu}(x)$, a basis of the tangent bundle $T\mathcal{M}$. Its dual is denoted as $\mathbf{e}^{a}(x)= e^{a}_{\mu} (x) dx^{\mu}$, being:
\begin{equation}
e^{a}_{\mu}(x) e_{b}^{\mu}(x) = \delta^{a}_{b}.
\end{equation} 
The tetrad field is related to the metric structure of the manifold by the orthonormal relation:

\begin{equation}
g_{\mu \nu}(x)=e^{a}_{\mu}(x)e^{b}_{\nu}(x) \eta_{ab},
\end{equation}
where $\eta_{ab}=\rm diag (-1,1,1,1)$ is the Minkowski metric in Cartesian coordinates. In this work, we use Greek letters $\mu,\nu,..= 0,..,3$ for space-time coordinate indices, and Latin letters $a,b,..= (0),..,(3)$ for Lorentzian tangent-space indices. The transformations that preserves orthonormality are the point-dependent (local) Lorentz transformations:
\begin{equation}
\mathbf{e}_{b'} (x)= \Lambda^{\; a}_{b'}(x) \mathbf{e}_{a}(x).
\end{equation}

Given a space-time $\mathcal{ST}=(\mathcal{M},g_{\mu \nu})$ it is possible to construct an affine geometry introducing a linear connection. GR$_{||}$ is based on the Weitzenb\"ock geometry \cite{Fecko:2006zy}, defined by the Weitzenb\"ock connection:
\begin{equation}
^{*} \Gamma^{\rho}_{\mu \nu}:= e^{\rho}_{a} \partial_{\nu} e^{a}_{\mu}.
\label{eq: Wei}
\end{equation}

The Weitzenb\"ock geometry is characterized by a null curvature tensor and a non trivial torsion tensor:
\begin{equation}
T^{\rho}_{\; \; \mu \nu}:=  2 \; ^{*}\Gamma^{\lambda}_{[ \mu \nu ]} \equiv e^{\rho}_a \Big ( \partial_{\nu} e^a_{\mu} - \partial_{\mu} e^a_{\nu} \Big ),
\end{equation}

The torsion tensor is not a scalar quantity under Lorentz tranformations since:
\begin{equation}
T^{\mu}_{\nu \rho} [ \mathbf{e}_{a}]- T^{\mu}_{\nu \rho} [ \mathbf{e}_{b'} ]= e^{\mu}_a e^{b}_{\rho} \omega^{a}_{\nu b} - e^{b}_{\nu} e^{\mu}_{a} \omega^{a}_{\rho b},
\end{equation}
where $\omega^{a}_{\nu b}:= \Lambda^{a}_{c} \partial_{\nu} \Lambda^{c}_{b}$ is known as the flat Lorentz connection. The link between the Weitzenb\"ock geometry and the Riemannian geometry of GR--- based on the Levi-Civita connection $\Gamma^{\rho}_{\mu \nu}$--- is given by the contortion tensor $K^{\rho}_{\mu \nu}$
\begin{equation}
K^{\rho}_{\mu \nu}=  ^{*} \Gamma ^{\rho}_{\mu \nu} - \Gamma^{\rho}_{\mu \nu} \equiv \frac{1}{2} g^{\rho \sigma} \Big ( T_{\nu \rho \mu} + T_{\mu \sigma \nu} - T_{\sigma \mu \nu} \Big ).
\end{equation}

The Lagrangian of the theory is built with cuadratic scalars of the torsion tensor:
\begin{equation}
\mathcal{L}_{T} \equiv \kappa \mathbb{T} =  \kappa \Sigma^{\rho \mu \nu} T_{ \rho \mu \nu},
\end{equation}
where $\mathbb{T}:=\frac{1}{4} T^{\rho}_{\; \; \mu \nu} T^{\; \;  \mu \nu}_{\rho} +\frac{1}{2} T^{\rho}_{\; \; \mu \nu} T^{\nu \mu}_{\; \; \;  \rho} - T^{\rho}_{\; \; \mu \rho}T^{\nu \mu}_{\; \; \; \nu}$ is the \textit{torsion} scalar and
\begin{equation}
\Sigma^{\rho \mu \nu}:= \frac{1}{2} \Big ( K^{\mu \nu \rho} -g^{\rho \nu} T^{\sigma \mu}_{\; \; \; \sigma} + g^{\rho \mu} T^{\sigma \nu}_{\; \; \; \sigma} \Big ).
\label{eq: superpotential}
\end{equation}
is the \textit{superpotential}. We can decomposed this Lagrangian as:
\begin{equation}
\mathbb{T} \equiv  -R  -  2 \nabla^{\mu}  T^{\nu}_{\; \; \mu \nu},
\label{eq: riccitorsion}
\end{equation}
where $R$ is the Riemannian Ricci scalar. Since both scalars differ by a boundary term, the dynamic equations of GR$_{||}$ are equivalent to Einstein equations \cite{local}. The complete action of GR$_{||}$ is then:
\begin{equation}
S[\mathbf{e}^{a}]= -\kappa \int  \mathbb{T} \: \sqrt{-g}  \: d^4x 	+ S_{M},
\end{equation}
and the field equations are:
\begin{equation}
\partial_{\nu} \Big ( \sqrt{-g} \Sigma^{a \lambda \nu} \Big ) = \frac{\sqrt{-g}}{4k} \Big ( t^{\lambda a} + \Theta^{\lambda a}  \Big ),
\label{eq: campo}
\end{equation}
where
\begin{equation}
t^{\lambda a}=  \kappa e^{a}_{\mu}  \Big ( 4 \Sigma^{bc \lambda} T_{bc}^{\; \; \mu} -g^{\lambda \mu} \Sigma^{bcd} T_{bcd} \Big ),
\label{eq: energymomentumgrav}
\end{equation}
is the energy-momentum tensor of gravity and $\delta\mathcal{L}_M/ \delta e_{a\rho}:= \sqrt{-g} \: e^{a}_{\nu} \Theta^{\nu}_{\rho}$ the matter energy-momentum tensor. From the teleparallel field equations (\ref{eq: campo}), considering the asymmetry of the superpotential in the last two indices, we obtain the conservation law:
\begin{equation}
\nabla_{\mu} \Big (t^{\mu a} + \Theta^{\mu a} \Big )  \equiv \partial_{\mu} \Big [ \sqrt{-g} \Big (t^{\mu a} + \Theta^{\mu a} \Big ) \Big ] = 0.
\label{eq: conservacion}
\end{equation}

Although $t^{\mu \nu}$ is a well-behaved tensor, it is gauge-dependent because of the non-invariance of the torsion tensor under local Lorentz transformations. However, the degrees of freedom in this definition acquire a physical meaning when we adopt the usual representation of the tetrad frame as a reference system \cite{Felice:2010cra}. Given a congruence $\mathcal{C}$ over $\mathcal{ST}$, we represent a reference system $K$ as a pair $(\mathcal{C},\mathbf{e}^{a})$, where we identified the zero component of the tetrad as the four-velocity of the curves in each point, $\mathbf{e}^{(0)}\equiv \mathbf{U}$, and $\mathbf{e}^{(i)}$ as the spatial frame.

The inertial properties of the frame on a given space-time can be characterized by the projected components of the contorsion tensor, i.e. the Ricci coefficients, through the relation:
\begin{equation}
e^{\nu}_c \nabla_{\nu} e^{a}_{\mu} = K^{a}_{bc} e^{b}_{\mu}.
\end{equation}

For instance, the tetrad frame changes through its time-like ($c=(0)$) component as
\begin{equation}
	U^{\nu} \nabla_{\nu} e^{a}_{\mu} = K^{a}_{b (0)} e^{b}_{\mu} = \phi^{a}_{b} e^{b}_{\mu},
\end{equation}
where $\phi_{ab}$ it is the so-called acceleration tensor  \cite{Mashhoon:2012az}, being $a_{(i)}:=\phi_{(0)(i)}$ the translational acceleration and $\Omega_{(i)}:=\epsilon_{(i)(j)(k)} \phi^{(j)(k)}$ the rotational acceleration (with respect of a Fermi-Walker transported frame). The values of a property represented by a tensor field $\mathbf{P}$ in a reference frame $K$ are then obtained from the projected tensor on the tetrad field, $ \mathbf{P}\cdot \mathbf{e} \: \hat{=} \: \mathcal{P}_x (K)$. The conserved four-momentum of space-time over an 3-hypersurface $\mathcal{D}$ in a reference system $K$ is then:
\begin{equation}
P^{a}= \int_{\mathcal{D}} (\Theta^{\mu \nu} + t^{\mu \nu} ) \: e_{\nu}^{a} \: d^{3}\mathcal{D}_{\mu},
\end{equation}
which depends intrinsically on the frame $K$ but is coordinate invariant. Due to the antisymmetry of the superpotential, in a compact region of space-time---if the superpotential is smooth--- we can use Gauss's law to express the total energy as: 
\begin{equation}
P^{a}= 2 \kappa \oint_{\partial \mathcal{D}} \Sigma^{a \mu \nu} d\mathcal{S}_{\mu \nu} = 4 \kappa \oint_{\partial \mathcal{D}} \Sigma^{a 01 } \sqrt{-g} d^2 \theta 
\label{eq: fourmomentum}
\end{equation}
where $\Sigma^{a \mu \nu} d\mathcal{S}_{\mu \nu}=2 \Sigma^{a01} \sqrt{-g} d^2 \theta$ in spherical-type coordinates (see Ref. \cite{poisson2004relativist}). The conservation law for $P^a$ can be written in a non-covariant---though practical--- form as 
\begin{equation}
\frac{dP^{a}}{dt}= \Phi_{M}^{a}  + \Phi^{a}_{G},
\label{eq: conserv}
\end{equation}
where
\begin{equation}
\Phi_{M}^{a}:= -\oint_{\partial D} \Theta^{ja} \: \sqrt{-g} \: dS_{j}, \quad \Phi_{G}^{a}:= -\oint_{\partial D} t^{ja} \: \sqrt{-g} \:  dS_{j},
\label{eq: flux}
\end{equation}
are the matter and gravitational fluxes, respectively \cite{maluf2007regularized}.

\section{Energy of a charged black hole}

The Reissner-Nordstrom geometry outside the source is given in Schwarzschild-type coordinates by

\begin{equation}
ds^2= -\alpha^2 (r) dt^2 + \frac{1}{\alpha^2 (r)} dr^2 + r^2 d\Omega,
\end{equation}
with $\alpha(r):= \sqrt{1-\frac{2M}{r} + \frac{Q^2}{r^2}}$ in geometrical units. The metric has two horizons in $r_{\pm}= M \pm \sqrt{M^2-Q^2}$, being $r_{+}$ an event horizon and $r_{-}$ a Cauchy horizon.

In order to obtain the energy distribution of the gravitational and electromagnetic field, first we must choose a reference system. We will begin with a stationary frame, a congruence with velocity $e_{(0)}^{\mu}= 1/\alpha(r) \delta^{\mu}_{0}$. In spherical coordinate, this frame is given by\footnote{It easy to check that this frame is holonomous, i.e. Cartesian, at infinity.}

\begin{equation}
e^{\: \mu}_{a} = \left ( \begin{array}{cccc}
\alpha(r)^{-1} & 0 & 0 & 0 \\
0 & \alpha(r) \cos\phi\sin\theta & \frac{\cos\theta \cos\phi  }{r} & -\frac{ \rm{csc} \theta
	\sin\phi}{r} \\
0 & \alpha(r) \sin\theta\sin\phi & \frac{\cos\theta \sin\phi}{r} & \frac{\cos\phi \rm{csc}\theta}{r} \\
0 & \alpha(r) \cos\theta & -\frac{\sin\theta}{r} & 0  \end{array} \right).
\label{eq: framestationary}
\end{equation}

The frame has zero rotation $\phi_{(i)(j)}=0$ and a radial acceleration given by:
\begin{equation}
\mathbf{a}=  \frac{-Q^2+M r}{r^3 \alpha(r)} \mathbf{r},
\end{equation}
where $\mathbf{r}:=( \cos \phi \sin \theta,  \sin \phi \sin \theta, \cos \theta)$. This is necessary to maintain the system in the stationary regime. The total energy contained in a 2-sphere of radio $R>r_{+}$ around the singularity for a constant time slice can be obtained from definition (\ref{eq: fourmomentum}):

\begin{equation}
P^{(0)}= 4 \kappa \int \nabla_{\mu} \Sigma^{(0) \nu \mu} d^3 \mathcal{D}_{\nu}
\label{eq: volumeint}
\end{equation}

Since the superpotential is singular in $r=0$, in order to use the Gauss theorem we must enclosed the singularity with a sphere $\epsilon<r_{-}$ and then take the limit $\epsilon \rightarrow 0$. Hence, using that $\Sigma^{(0)0 i}=(\frac{1-\alpha}{r},0,0)$ in Schwarzchild-type coordinates, integral (\ref{eq: volumeint}) for a spherical shell with interior radius $\epsilon$ and external radius $R$ can be converted into a surface integral as
\begin{equation*}
P^{(0)}_{(\epsilon, R)}= \Big ( \oint_{ \mathcal{S}(R)}-\oint_{ \mathcal{S}(\epsilon)} \Big ) 4  \kappa  \Sigma^{(0) 0 1} \sqrt{-g} d\theta d\phi = R \Big ( 1- \alpha(R) \Big) - \epsilon \Big ( 1- \alpha(\epsilon) \Big),
\end{equation*}

Taking the limit $\epsilon\rightarrow0$, we get
\begin{equation*}
P^{(0)}(R)= R \Big ( 1- \alpha(R) \Big) + |Q|,
\label{eq: energíarn}
\end{equation*}

The first term of $P^{(0)}$ is the quasi-local Brown-York energy, obtained directly from a surface integral \cite{Lundgren:2006fu}. In Ref. \cite{Castello} the authors have found this Brown-York energy using GR$_{||}$ and applied it to analyze the thermodynamics of the event horizon. However, this result is not entirely correct. Since the teleparallel energy is defined through a volume integral, the correct result must take into account the singularities into the surface integral. In such a case, the module of the charge appears as an extra contribution to the energy. We will show this explicitly calculating the volume integral. This is a general feature of the pseudo-tensor approach to obtain the gravitational energy that is often unnoticed in the literature. Usually, the extra contribution of the singularity is zero. For instance, in Schwarzchild space-time, where $Q=0$, we obtain exactly the quasi-local Brown-York mass.

The total energy is not defined in $r_{-}<R<r_{+}$ since no stationary frame exists in this region. The energy is always positive and taking the asymptotic limit gives
\begin{equation}
P^{(0)}(R\rightarrow\infty)= M+|Q|.
\end{equation}

This is an undesired result since the energy is not asymptotically the ADM mass. As we said, the extra term appears because of the character of the singularity in the RN space-time. The Brown-York energy of the RN space-time is negative inside the inner-horizon, yielding $-|Q|$ in $R \rightarrow 0$. In GR$_{||}$, the singularity has zero energy. 

To the next order for large $R$, we get
\begin{equation}
P^{(0)}\approx M + |Q| +\frac{M^2}{2 R} - \frac{Q^2}{2 R} + \mathcal{O}(R^{-3}),
\end{equation}
where equal contributions of the mass and charge appears. The spatial momentum of the field is zero, $P^{(i)}=0$ as expected for an stationary frame in an stationary space-time. 

Now, we calculate the energy-momentum tensors to obtain each energy separately. From definition (\ref{eq: energymomentumgrav}) and after straightforward algebraic steps, we obtain

\begin{equation}
t^{\mu \nu} = \left(
\begin{array}{cccc}
-\frac{(1-\alpha(r))^2}{8 \pi  r^2 \alpha(r)^2} & 0 & 0 & 0 \\
0 & \frac{\sigma(r)}{8 \pi r^6}& 0 & 0 \\
0 & 0 & \frac{(\alpha(r)-1)}{\alpha(r)}\frac{-Q^2+M r}{8 \pi  r^6} & 0 \\
0 & 0 & 0 & \frac{(\alpha(r)-1)}{\alpha(r)}\frac{(-Q^2+M r)\csc^2\theta}{8 \pi  r^6} \\
\end{array}
\right),
\label{eq: tenet}
\end{equation}
where we define $\sigma(r):=-r^2 Q^2 \alpha(r)^2- 2(1-\alpha(r))r^4 + M(-2+\alpha(r))$. For the electromagnetic field we get
\begin{equation}
\Theta^{\mu \nu} = \left(
\begin{array}{cccc}
\frac{Q^2}{8 \pi  r^4 \alpha(r)^2} & 0 & 0 & 0 \\
0 & -\frac{Q^2 \alpha(r)^2}{8 \pi  r^4} & 0 & 0 \\
0 & 0 & \frac{Q^2}{8 \pi  r^6} & 0 \\
0 & 0 & 0 & \frac{Q^2 \rm{csc}^2\theta}{8 \pi  r^6} \\
\end{array}
\right).
\label{eq: tenem}
\end{equation}

In the asymptotic limit, both tensors behave in the same way
\begin{equation}
t^{11} \approx -\frac{M^2}{8 \pi r^4}, \quad t^{22} \approx \frac{M^2}{8 \pi r^4},
\end{equation}
\begin{equation}
\Theta^{11}\approx\frac{Q^2}{8 \pi r^4}, \quad \Theta^{22}\approx-\frac{Q^2}{8 \pi r^4}, 
\end{equation}
in agreement with the gravitoelectric behaviour of the field in this region \cite{Costa:2012cw}. On the other hand, the energy density---\textit{measure} in the stationary frame--- is given by projecting the energy momentum tensor on the tetrad:
\begin{equation}
\Theta^{(1)(1)} = \frac{Q^2}{8 \pi r^4},
\end{equation}
\begin{equation}
t^{(1)(1)} = -\frac{M^2}{8 \pi r^4} - \frac{M^3}{8 \pi r^5}+\frac{M}{r} \Theta^{(1)(1)}+\mathcal{O}(r^{-6}),
\end{equation}
where electromagnetic contribution appears in the gravitational energy density in the next order. It is interesting to notice that in the case of an extreme charged black hole, where $M^2=Q^2$, the energy-momentum tensors cancel each other in the $r>2M$ region. The total energy, given by $P^{0}=2R$, is totally enclosed inside the horizon. 

From (\ref{eq: fourmomentum}) we can calculate the energy contribution of each field inside a sphere of radius $R$, obtaining:
\begin{equation}
P^{(0)}_{EM}= \int drd\theta d\phi \sqrt{-g} \Theta^{0 (0)} = \int drd\theta d\phi \sqrt{-g} \frac{Q^2}{8 \pi  r^4 \alpha(r)}
\end{equation}
\begin{equation*}
=\frac{Q}{2} \log \Big[ \frac{r}{Mr-Q(Q+r\alpha(r))} \Big] \Big|^{r=R}_{r=0},
\end{equation*}
and
\begin{equation}
P^{(0)}_{G}= \int drd\theta d\phi  \sqrt{-g} t^{0 (0)} = - \int drd\theta d\phi  \sqrt{-g} \frac{(1-\alpha(r))^2}{8 \pi r^2 \alpha(r)},
\end{equation}
\begin{equation*}
= R \Big(1-\alpha(R)\Big)+|Q|-\frac{Q}{2} \log \Big[ \frac{r}{Mr-Q(Q+r\alpha(r))} \Big] \Big|^{r=R}_{r=0}.
\end{equation*}

Both expressions have self-energy problems and diverge in $r=0$, as it is expected for a point-charged particle. However, the sum is finite, yielding $P^{(0)} = P^{(0)}_{G} + P^{(0)}_{EM}$. This explicit self-regularization of the total energy due to the gravitational interaction have already been commented in the literature \cite{Lundgren:2006fu} and it seems a general feature of gravity. If we calculate both energies contained between a sphere of radius $R$ and infinity we obtain
\begin{equation}
P^{(0)}(R,\infty)_{G}=- R \Big(1-\alpha(R)\Big)+\frac{Q}{2} \log \Big[ \frac{r}{Mr-Q(Q+r\alpha(r))} \Big] - \frac{1}{2}Q \log(M-Q) + M,
\end{equation}
\begin{equation}
P^{(0)}(R,\infty)_{EM}=-\frac{Q}{2} \log \Big[ \frac{r}{Mr-Q(Q+r\alpha(r))} \Big] + \frac{1}{2}Q \log(M-Q).
\end{equation}

In the asymptotic limits, these results reduce to the expected electroestatic and gravitoelectroestatic behavior
\begin{equation}
P^{(0)}(R,\infty)_{G}\approx -\frac{M^2}{2R},
\end{equation}
\begin{equation}
P^{(0)}(R,\infty)_{EM}\approx \frac{Q^2}{2R}.
\end{equation}
for $R \rightarrow \infty$.

\section{Accelerated observers and radiation}

The energy in the teleparallel approach depends on the chosen frame. In order to analyze further  the differences between the  electromagnetic field and gravity, let us consider a frame accelerated in the $\mathbf{e}_{(1)}$ direction of the stationary frame. This amounts to transform the tetrad with a time dependent Lorentz matrix (see Ref. \cite{Maluf:2010fb})

\begin{equation}
\Lambda_{a'}^{b}= \left(
\begin{array}{cccc}
\gamma & \beta \gamma & 0 & 0 \\
\beta \gamma & \gamma & 0 & 0 \\
0 & 0 & 1 & 0 \\
0 & 0 & 0 & 1 \\
\end{array}
\right),
\end{equation}
where $\gamma:= \frac{1}{\sqrt{1-\beta^2}}$, and $\beta=\beta(t)$ is the time dependent velocity of the frame (at infinity). Applying this matrix to the stationary frame (\ref{eq: framestationary}), we obtain:

\begin{equation}
e^{\mu}_{a} = \left ( \begin{array}{cccc}
\alpha(r)^{-1} \gamma & \gamma \beta \alpha(r) \cos\phi \sin\theta & \gamma \beta \frac{\cos\theta \cos\phi  }{r} & - \gamma \beta \frac{ \rm{csc} \theta
	\sin\phi}{r} \\
 \gamma \beta \alpha(r)\beta & \gamma \alpha(r) \cos\phi\sin\theta & \gamma \frac{\cos\theta \cos\phi  }{r} & - \gamma \frac{ \rm{csc} \theta
	\sin\phi}{r} \\
0 & \alpha(r) \sin\theta\sin\phi & \frac{\cos\theta \sin\phi}{r} & \frac{\cos\phi \rm{csc}\theta}{r} \\
0 & \alpha(r) \cos\theta & -\frac{\sin\theta}{r} & 0  \end{array} \right).
\label{eq: accframe}
\end{equation}

The frame acquires an additional acceleration besides the inertial-static acceleration:

\begin{equation}
a^{(1)}= \gamma \cos \phi \sin \theta \frac{-Q^2+M r}{r^3 \alpha(r)} + \gamma^3 \frac{\beta'(t)}{\alpha(r)},
\end{equation}
\begin{equation*}
a^{(2)}= \gamma^2 \sin \phi \sin \theta \frac{-Q^2+M r}{r^3 \alpha(r)} + 
\beta^2 \gamma^2 \sin \phi \sin \theta \frac{-Q^2+M r+ r^2(1-\alpha(r))}{r^3 \alpha(r)},
\end{equation*}
\begin{equation*}
a^{(3)}= \gamma^2 \cos\theta \frac{-Q^2+M r}{r^3 \alpha(r)} + 
\beta^2 \gamma^2 \cos\theta \frac{-Q^2+2M r- r^2(1-\alpha(r))}{r^3 \alpha(r)},
\end{equation*}
and also a rotation induced by the boost ( $\phi_{(i)(j)}\neq 0$). From the superpotential, we obtain a total momentum of the field given by
\begin{equation}
\tilde{P}^{a}=(\gamma P^{(0)}, \beta \gamma P^{(0)},0,0),
\end{equation}
where $P^{(0)}$ is the energy of the field in the stationary frame. We see then that the four-momentum of the field behaves as the four-momentum of a particle in special relativity. This result was obtained by Maluf \cite{Maluf:2004vc} for a Schwarzschild black hole in the asymptotic limit using isotropic coordinates. Here, we have shown that this result holds everywhere for the RN space-time. 

If $\beta'(t)\neq 0$, then the leading order of the gravitational energy density is quadratic:

\begin{equation}
t^{(1)(1)}= \frac{M^2}{4 \pi r^2} \cos\phi \sin\theta \beta(t)^2 \gamma^4 \beta'(t),
\label{eq: accgrav}
\end{equation}
while the energy-momentum tensor of the electromagnetic field remains of order $\mathcal{O}(r^{-4})$:
\begin{equation}
\Theta^{(1)(1)}= \frac{Q^2}{4 \pi r^4} \gamma^2 \Big [ -1 + (-\cos^2\theta + \cos2 \phi \sin^2\theta)  \beta(t)^2 \Big ] .
\label{eq: accmatt}
\end{equation}

The similarities between both fields, holding in the stationary frame, are lost when the frame is accelerated at infinity. Indeed, from (\ref{eq: accgrav})  and (\ref{eq: accmatt}), we see that an observer in the accelerated frame will detect gravitational radiation but no electromagnetic radiation. To see this explicitly, let us now calculate the energy fluxes. From the conservation law (\ref{eq: conserv}) and (\ref{eq: flux}), we get that the total variation of the energy is:
\begin{equation}
\frac{d\tilde{P}^{(0)}}{dt}= P^{(0)} \beta \gamma^3 \beta'(t) = \Phi^{(0)}_{G}+\Phi^{(0)}_{EM}.
\label{eq: conserflu}
\end{equation}

To see the contribution of each field to (\ref{eq: conserflu}) we calculate each flux separately. From (\ref{eq: flux}) we obtain for the electromagnetic field:
\begin{equation}
\Phi^{(0)}_{EM} = \oint \Theta^{(0)1} \sqrt{-g} d\theta d\phi = \oint \frac{Q^2}{8 \pi r^4} \alpha(r) \gamma \beta(t) \cos\phi \sin^2\theta  r^2 d\theta d\phi = 0.
\end{equation}

There is no radiation due to the electromagnetic field in this frame. For the gravitational field, using the same procedure as Section 3, we obtain:
\begin{equation}
\oint t^{(0)1} \sqrt{-g} d\theta d\phi =
\end{equation}
\begin{equation*}
=\oint \Big [ \frac{\cos\phi \sin\theta}{\gamma^2} \xi(r) +2r^3 (1-\alpha(r))\beta'(t)) \Big] \frac{\gamma^3 \beta(t)}{8 \pi r^4}  r^2 \sin\theta d\theta d\phi,
\end{equation*}
where $\xi(r)=2Mr-2r^2+\alpha(r)(2r^2-Q^2)$. Integrating we get
\begin{equation}
\oint t^{(0)1} \sqrt{-g} d\theta d\phi =R(1-\alpha(R)) \beta \gamma^3 \beta'(t).
\end{equation}
Considering the extra contribution of the singularity, we obtain:
\begin{equation}
\Phi^{(0)}_{G} = P^{(0)} \beta \gamma^3 \beta'(t).
\end{equation}

These results agree with those obtained in \cite{Maluf:2004vc} for a Schwarzschild black hole and in \cite{Maluf:2010fb} for a charged particle in Minkowski space-time. The latter seems to contradict the result by Rohrlich \cite{rohrlich2007classical}, who found that an static charged-particle would radiate from the point of view of an accelerated observer. The inequivalence between these results could be resolved by further analysis of our tetrad accelerated frame in relation to the coordinate approach used in \cite{rohrlich2007classical}. 

The frame (\ref{eq: accframe}) is interesting in its own right. In flat space-time, it easy to show that the frame $e_{\: \mu}^{a}$ has a global acceleration given by
\begin{equation}
\phi_{(0)(1)}= \frac{d(\beta \gamma)}{dt} = a,
\end{equation}
which is the well-known hyperbolic acceleration for constant $a$. However, the metric is still Minkowskian:
\begin{equation}
e^a_{\mu}(t)\eta_{ab} e^{b}_{\nu}(t)= \eta_{\mu \nu},
\end{equation}
i.e. no horizon arises for this observer. This is an interesting difference with Rindler's frame which we will investigate elsewhere.

\section{Conclusions}

We have studied the energy distribution of a charged black hole using the Teleparallel framework of General Relativity. Adopting the usual representation of tetrad fields as reference frames, we have calculated the gravitational plus matter energy measured by a stationary frame and by an accelerated frame. Since the theory admits a well-behaved energy-momentum tensor for gravity, we obtained the contribution of each field (electromagnetic and gravity) to the total energy. We showed that separately both energies diverge---as it should be the case for a point charged particle--- but the total energy is finite. We have proved that the teleparallel energy is not totally equivalent to a quasilocal energy definition since we have a non-null contribution to the volume integral in the singularity.

In the accelerated frame, we obtained that both fields behave very differently, even in the asymptotic (Newtonian) limit. In this frame, a Poynting-like flux due to the gravitational field appears, while such effects are absent for the electromagnetic field. This means that in this frame, some kind of gravitational radiation appears though space-time is stationary. Since in this frame the curvature is time dependent, $R^{a}_{\: bcd}(t)= R^{\mu}_{\: \nu \rho \sigma} e^{a}_{\: \mu}(t) e^{\nu}_{\: b}(t) e^{\rho}_{\: c}(t) e^{\sigma}_{\: d}(t)$, the observers will detect nearby deviations that are time dependent. Better understanding of the asymmetry between both fields in this case would require to investigate the relation of the inertial properties of the observer and the gravitational energy concept. 

The approach we have developed here can be useful to investigate the energy exchange between matter and gravity in a non stationary space-time such as those of a Petrov D type and cosmological scenarios. Such studies will be presented elsewhere.

\acknowledgments
This work was supported by grants PICT 2012-00878 (Agencia Nacional de Promoci\'on Cient\'ifica y Tecnol\'ogica, Argentina) and AYA 2013-47447-C3-1-P (Ministro de Educaci\'on, Cultura y Deporte, Espa\~na). We would like to thank Federico Lopez Armengol for helpful comments

\bibliographystyle{ieeetr}
\bibliography{bibliography}

\end{document}